# Electrical conductivity of vanadium dioxide switching channel

**A. Pergament** [*], **P. Boriskov, N. Kuldin, and A. Velichko**

Physics and Technology Department, Petrozavodsk State University, Lenin Ave 33, 185910 Petrozavodsk, Russian Federation

[*] Corresponding author: e-mail aperg@psu.karelia.ru, Phone: +07 8142 719 658, Fax: +07 8142 711 000

The electrical conductivity of the switching channel of vanadium dioxide thin-film sandwich structures is studied over a wide temperature range (15 – 300 K). It is shown that the electrical resistance of the channel varies with temperature as $R \sim \exp(aT - b/T)$ in the high-temperature region (above 70 K). The experimental results are discussed from the viewpoint of the small polaron hopping conduction theory which takes into account the influence of thermal lattice vibrations onto the resonance integral.

**1 Introduction** The small polaron theory, taking into account the effect of thermal lattice vibrations on the resonance integral, has been developed in [1]. From the viewpoint of experiment, the most important result of this theory is that the hopping conduction mechanism is essentially modified as compared to the standard temperature dependence of conductivity $\sigma \sim \exp(-E_a/k_B T)$, that is, $\ln(\sigma) \sim 1/T$, with the activation energy $E_a$. In particular, if the mean-square thermal displacement $\langle \rho^2 \rangle$ of atoms is of the order of (or greater than) the squared polaron localization radius $R_p$, then, at high temperatures, the temperature dependence of conductivity becomes

$$\ln(\sigma) \sim T. \qquad (1)$$

Generally, the conductivity can be written as [1, 2]:

$$\sigma = en \frac{ea^2}{4\pi^{1/2}\hbar} \frac{I^2}{E_a^{1/2}(k_B T)^{3/2}} \exp(-E_a/k_B T + k_B T/\varepsilon), \qquad (2)$$

where $a$ is the interatomic distance, $e$ – the electron charge, $I$ – the resonance integral, $\hbar$ and $k_B$ – the Planck and Boltzmann constants, respectively. The constant $\varepsilon$ is independent of temperature and proportional to the squared polaron radius. In the high-temperature region (when $2k_B T > \hbar\omega_o$, where $\omega_o$ is the frequency of an optical phonon), this constant is given by the following expression [2]:

$$\varepsilon = \tfrac{1}{4} M\omega^2 R_p^2. \qquad (3)$$

Here $M$ is the mass of an atom and $\omega$ is a characteristic phonon frequency.

The theory [1] takes into account the influence of the lattice atom thermal displacements upon the small polaron inter-site hopping probability. The displacements of atoms lead to the change in the neighbor site wave functions overlap, and the latter contributes to the resonance integral $I$. The value of $I$ depends on the inter-site hopping distance $R$ as: $I \sim \exp(-R/R_p)$. The hopping mobility, in turn, is proportional to $I^2$. More detailed derivations of Equations (2) and (3), as well as the model validation in some specific cases, one can find in the works [1-3]. In the recent works [2, 4-6], it has been shown that many transition metal compounds, exhibiting the metal-insulator transition (MIT), obey the law (1), and the values of $R_p$ have been calculated from Eq. (3). The materials tested were: $VO_2$ (the MIT temperature $T_t = 340$ K) [2], $CuIr_2S_4$ ($T_t = 220$ K) [4], $V_2O_3$ ($T_t = 150$ K) [5], and $V_4O_7$ ($T_t = 240$ K) [6]. Earlier we had reported on the switching effect in vanadium dioxide thin-film sandwich structures [7]. The switching parameters (threshold voltage $V_{th}$ and other) were measured in a wide temperature range (15−340 K), and the switching mechanism based on the Mott MIT occurring in the electric field was proposed [7-9].

In this work we report on the temperature dependence of the $VO_2$ switching channel electrical conductivity. The results are analyzed in light of the above described model and discussed in comparison with the data for vanadium dioxide single crystals.

**2 Experimental** The sandwich devices under study were fabricated by anodic oxidation of vanadium metal substrates [9], and Au electrodes were thermally evaporated at room temperature onto the surfaces of the films to complete the metal-oxide-metal (MOM) structures. The vanadium oxide film thickness was 180 nm.

The switching channel, consisting of $VO_2$, forms during electroforming, and the current-voltage (I–V) characteristic becomes S-shaped [7-9]. The switching effect is conditioned by the development of an electrothermal insta-bility in this channel. When a voltage is applied, the channel is heated up to $T = T_t$ at $V = V_{th}$, and the structure undergoes a transition from an OFF insulating state to an ON metallic state. In high electric fields (~ $10^6$ V/cm), non-thermal electronic effects contribute to the Mott MIT in $VO_2$ and, thereby, modify the switching mechanism.

The *I-V* characteristics of the electroformed MOM structures were studied by a two-probe method [7, 8], and the OFF-state resistance $R_{OFF}$ was measured near zero bias ($V \ll V_{th}$). Temperature dependences of $R_{OFF}$ were recorded using a Gifford-McMahon cycle cryorefrigerator. At low temperatures, when $R_{OFF}$ was as high as $10^8 – 10^9$ Ω, the current was measured with a picoammeter TESLA BM-545.

**3 Results and discussion** Figure 1 shows the dependence of the resistance $R_{OFF}$ on temperature. One can see that the curves are nonlinear in Arrhenius coordinates. Overall these data better fit a relationship of Eq. (1) – see Fig. 2. In spite of a satisfactory linearization, the slope *B* may scarcely be used for the calculation of the constant $\varepsilon$, though formally, in Eq. (2), $B = k_B/\varepsilon$. The point is that the condition of high temperature is obviously not fulfilled in the entire temperature range, and the term $E_a/k_BT$ is not negligibly small.

Note that Eq. (2) can be rewritten as:

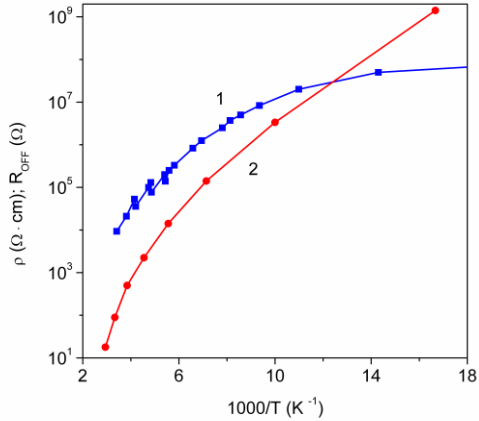

**Figure 1** $VO_2$ switching channel resistance $R_{OFF}$ (1) as a function of reciprocal temperature measured at cooling. Specific resistivity $\rho$ (2) of a vanadium dioxide single crystal [2] is shown for comparison.

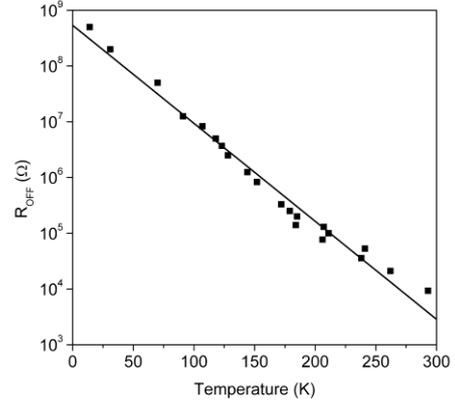

**Figure 2** Temperature dependence of $R_{OFF}$ in semi-logarithmic coordinates. Solid line indicates a linear fit $\ln(R_{OFF}) = A - BT$ (where $B = 0.0405$ K$^{-1}$) with the approximation reliability of 0.98. The data scattering is due to multiple measurements at cooling and heating.

$$\ln(\sigma T^{3/2}) \cdot T = CT - \frac{E_a}{k_B} + \frac{k_B}{\varepsilon} T^2, \quad (4)$$

where *C* is a constant which does not depend on temperature. Thus, when plotting $F \equiv \ln(T^{3/2}/R_{OFF}) \cdot T$ against *T*, the ensuing graph should be a parabola. These data are presented in Fig. 3(a). The approximation reliability in this case is rather low (0.93), but we could, at least, obtain the value of $E_a/k_B$ (see the figure caption), and then apply a procedure suggested in the works [2, 6], *viz.* plot the value of $\ln(\sigma T^{3/2}) + E_a/k_BT$ (or $-[F/T + E_a/k_BT]$, Fig.3(b)) versus *T*. This curve just gives the refined coefficient $B = k_B/\varepsilon$ which turns out to be equal to 0.04 K$^{-1}$.

It should be noted that the analysis of the data on the temperature dependence of conductivity suggested in the works [2, 6] is somewhat tangled; the authors re-arrange Eq. (2) in such a way:

$$\ln(\sigma T^{3/2}) - \frac{k_B}{\varepsilon} T = A - \frac{E_a}{k_B T}, \quad (5\text{-a})$$

$$\ln(\sigma T^{3/2}) + \frac{E_a}{k_B T} = A + \frac{k_B}{\varepsilon} T, \quad (5\text{-b})$$

and then adjust the values of $E_a$ and $\varepsilon$ to the quantities which result in the best approximations of both Equations (5) by linear relationships. Figure 4 represents the experimental results of the work [2] processed as suggested in the present work. Curve equations and the approximation reliability factors are inserted into the figures. The slope $B = k_B/\varepsilon$ in Fig. 4(b) is equal to ~0.035 K$^{-1}$ which is close to the result for the $VO_2$ switching channel (0.040 K$^{-1}$ – see above).

From the data of Fig. 3(b), one can calculate the value of $\varepsilon$ and estimate the polaron localization radius using Eq. (3).

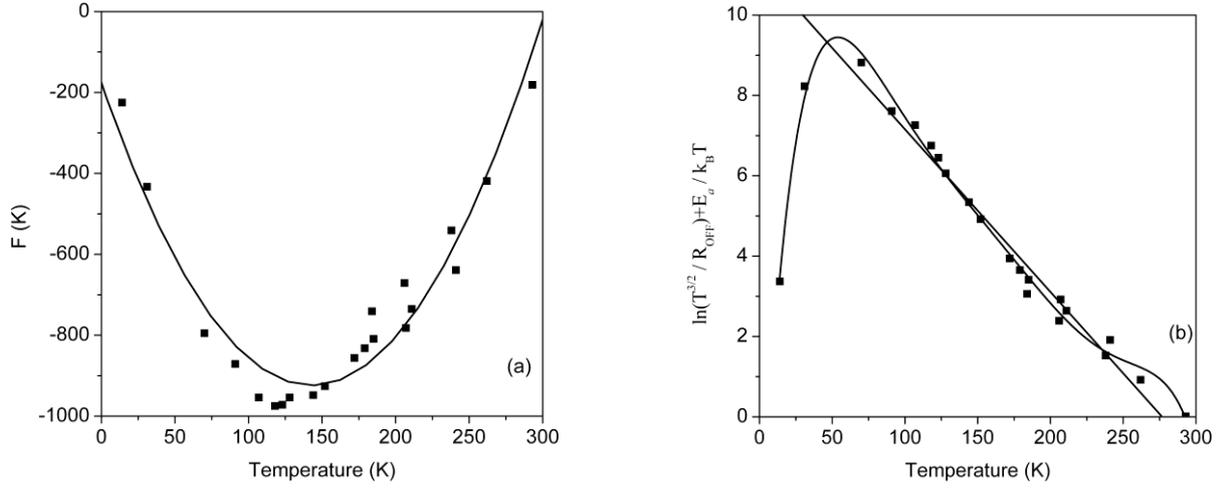

**Figure 3** (a) Parabolic relationship between the parameter $F = \ln(T^{3/2}/R_{OFF}) \cdot T$ and temperature. The equation is $F = 0.037T^2 - 10.45T - 177.8$ with the approximation reliability of 0.93. (b) The dependence of the parameter $-(F/T + 177.8/T)$ on temperature. Linear fit in the temperature range 70 – 290 K gives the slope $B = k_B/\varepsilon = 0.040$ K$^{-1}$ with the approximation reliability of 0.98.

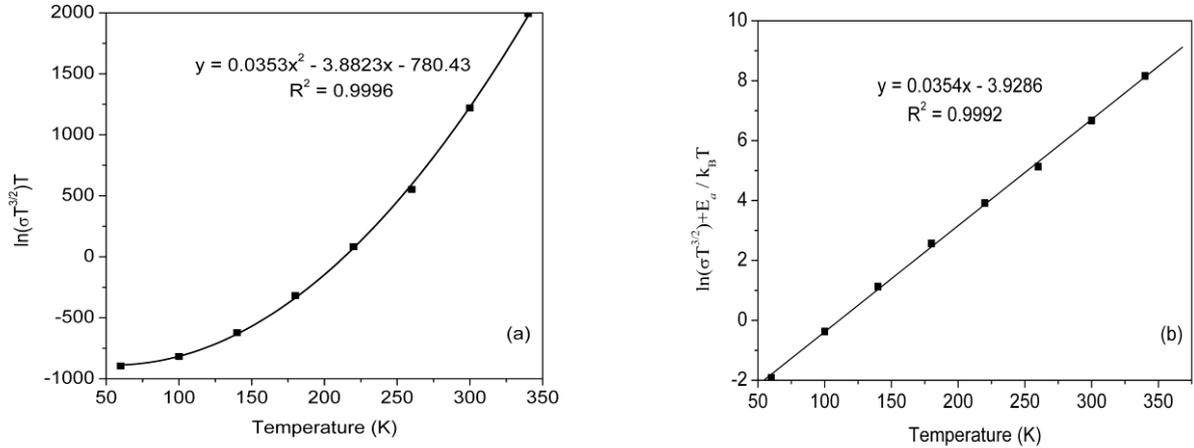

**Figure 4** Data on electrical conductivity of VO$_2$ single crystal (see Fig. 1, curve 2) processed in the same way as in Fig. 3 (a, b).

The results are as follows: $\varepsilon = 2.16 \times 10^{-3}$ eV (in [2] $\varepsilon = 2.46 \times 10^{-3}$ eV), and $R_p \approx 0.5 \times 10^{-11}$ m. The values of $M = 8.5 \times 10^{-26}$ kg (the mass of a vanadium atom) and $\omega = \omega_0 = 2.6 \times 10^{13}$ s$^{-1}$ (a low-frequency optical phonon mode of VO$_2$ [2, 10]) were used for the calculations. Such a small localization radius (~0.05 Å) is evidently underestimated. Discussing this result, the authors of the work [2] do not give any explanation except for obscure speculations that "this small value of $R_p$ should be considered merely as an effective localization radius".

If however we calculate, conversely, the value of $\omega$ from Eq. (3) presuming that $R_p$ is equal to a reasonable estimate, e.g., to the correlation length $\xi$ for the semiconducting phase of VO$_2$ ($\xi = 1-2$ Å [11]), we obtain $\omega \sim$ $(0.6–1.3) \times 10^{12}$ s$^{-1}$. Such a frequency corresponds to either an acoustic phonon, or a soft mode of optical phonons. The softening of the phonon modes at the MIT in VO$_2$ has been observed [10] and discussed with regard to the transition mechanism [12].

This assumption is also confirmed by the temperature-dependent sound velocity $v_s$ measurements. It has been shown [13] that, as the temperature increases, the value of $v_s$ (and the latter, as is known, is proportional to the Debye frequency [14]) decreases, and this decrease commences long before the transition to occur at $T_t = 340$ K (in the work [13], the data on $v_s$ are presented in the temperature range 240–350 K).

Apparently, the phonon mode softening, which inheres in any structural phase transition, contributes to the lowering of the characteristic frequency $\omega$ in Eq. (3). Note that the value of $\omega$ is not determined exactly in the model of the phonon-assisted small polaron hopping conduction [1, 3]. One can surmise that it could be a certain mean frequency constituting of a mixture of different (both acoustic and optical) phonon modes. Actually, Eq. (3) is a simplified expression, and an accurate equation for $\varepsilon$ contains a partition function all over the phonon spectrum [1]:

$$\varepsilon \sim \left( \sum_q \frac{1-\cos(\mathbf{q}\mathbf{g})}{M\omega^2(q)} \right)^{-1}, \quad (6)$$

where $\mathbf{q}$ is the phonon wavevector, and $\mathbf{g}$ is the radius-vector to the nearest neighbor site ($|\mathbf{g}| = a$).

The theory of small polaron transport, with its Eq. (2), is valid only provided that [1-4]

$$R_p^2 \geq \langle \rho^2 \rangle. \quad (7)$$

On the other hand, it is well known that any structural instability gives rise to a divergence of the atomic mean-square displacements [14]. Therefore, in the vicinity of the MIT, the value of $\rho = (\langle \rho^2 \rangle)^{1/2}$ might be ~ 1–2 Å = $\xi$, and not 0.1 Å, as supposed in [2]. Moreover, in the monograph [3], the ultra-small polaron radius is directly associated with the ionic radius. For $VO_2$ the radius of the vanadium $d$-shell (i.e. the ionic radius) is ~ 0.8 Å, and the atomic radius of vanadium equals to 1.32 Å [11, 15, 16]. Thus, the estimates of $R_p \sim 1$ Å and $\omega \sim 1.3 \times 10^{12}$ s$^{-1}$ seem to be quite reasonable.

**4 Conclusion** To summarize, it is shown that the temperature dependence of the $VO_2$ switching channel resistance is described in terms of the small polaron hopping conduction theory which takes into account the influence of thermal vibrations of atoms onto the resonance integral (i.e. the nearest sites wave-function overlap). Simple and fairly effective technique for processing of experimental data to fit Eq. (2) is proposed. This technique consists in plotting a graph of the value of $F$ (or $\ln(\sigma T^{3/2}) \cdot T$) as a function of $T$; such a graph should represent a square polynomial curve. The polynomial coefficients straight away give the unknown values of $E_a$ and $\varepsilon$. For rather imperfect polycrystalline vanadium dioxide in the MOM structure switching channel, $E_a = 15.3$ meV (see Fig. 3 – $E_a = 177.8/k_B$), while for $VO_2$ single crystals, $E_a = 67.3$ meV (from the graph of Fig. 4; in [2] $E_a = 66.5$ meV). Thus, the activation energy in such materials may not be determined simply from the slope of $\sigma(T)$ plotted in Arrhenius coordinates in a narrow temperature range, and the analysis described above, using Eq. (4), is required.

For a $VO_2$ switching channel, the $\varepsilon$ value has been found to be $2.16 \times 10^{-3}$ eV which is close to the result obtained for a $VO_2$ single crystal ($\varepsilon = 2.46 \times 10^{-3}$ eV) [2].

This means that the effect discussed is really local, because the second term in exponent of Eq. (2) is independent of the crystallite size (in contrast to the activation energy, which, as we saw above, does depend on whether we have a single crystal or a polycrystalline film). The polaron localization radius and characteristic phonon frequency have been estimated to be $R_p \sim 1$ Å and $\omega \sim 1.3 \times 10^{12}$ s$^{-1}$, respectively, and this conclusion is supported by the phenomenon of softening of the phonon modes and divergence of the atomic mean-square displacements (which ensures the fulfillment of the condition given by Eq. (7)). This phenomenon is inherent not only in the MIT in $VO_2$, but in any structural phase transition. That is why the temperature dependence of conductivity corresponding to Eq. (2) is observed mostly in the transition metal compounds undergoing the MIT [2, 4-6] (see also [3] and references therein).

Finally, it is believed that the model verification, i.e. the testing of conformity of the $\sigma(T)$ dependence with Eq. (2), would be of interest for other materials exhibiting MIT, especially for those with *re-entrant* MITs. In compounds exhibiting the inverse (or re-entrant) transitions [15], such as $NiS_{2-x}Se_x$ [17], nonstoichiometric EuO, and CMR-manganites [18], the semiconducting phase is high-temperature, and the condition $2k_BT > \hbar\omega_o$ is easier to fulfill. Note that the small polaron hopping conductivity in manganites has been recently reported, for example, in [19].

**Acknowledgements** This work was supported by the Ministry of Education and Science of Russian Federation through the "Development of Scientific Potential of High School" Program (projects No. 4978 and 8051), "Scientific and Educational Community of Innovation Russia" Program (projects No. P1156 and P1220) and the U.S. CRDF grant (award No. Y5-P-13-01). The authors are also indebted to V. N. Andreev for stimulating discussions.